# RAMAN SCATTERING IN IRON-BASED SUPERCONDUCTORS

A. M. ZHANG and Q. M. ZHANG[*]

*Department of Physics, Renmin University, Beijing 100872, P.R. China*

Iron-based superconducting layered compounds have the second highest transition temperature after cuprate superconductors. Their discovery is a milestone in the history of high-temperature superconductivity and will have profound implications for high-temperature superconducting mechanism as well as industrial applications. Raman scattering has been extensively applied to correlated electron systems including the new superconductors due to its unique ability to probe multiple primary excitations and their coupling. In this review, we will give a brief summary of the existing Raman experiments in the iron-based materials and their implication for pairing mechanism in particular. And we will also address some open issues from the experiments.

## 1. Background

Raman scattering, also known as inelastic light scattering, was discovered in 1928 by C. V. Raman. It can be used to detect many kinds of excitations in solids, such as phonons,[1] magnons,[2] excitons[3] etc. With the invention of laser and the continuous improvement of monochromator and new-generation detectors, Raman scattering has become a conventional and fundamental technique in many fields. Superconductivity is one of the most important frontiers in condensed matter physics and much Raman scattering work has been done in the field so far. Basically, Raman scattering can probe phonons and related electron-phonon coupling. The information on superconducting gap size can be drawn from cooper-pair-breaking peak in electronic Raman scattering spectra. Symmetry analysis is a unique advantage of Raman scattering. In a non-isotropic superconductor, the position, shape and low-energy behavior of pair-breaking peak will substantially change in different symmetry channels, which gives a unique way to explore pairing symmetry. Raman scattering has also been employed to detect two-magnon procedure in spin-ordered systems, from which we can obtain accurate exchange energies and learn the evolution of spin fluctuation. Actually, Raman scattering plays a crucial role in the study of cuprate superconductors.

Since its discovery in 2008, iron-based superconductor has attracted a lot of interests due to its second highest transition temperature after cuprate superconductor. So far, all known iron-based superconductors can be divided into six categories in structure: 1, "1111", represented by F-doped LaFeAsO, is the first iron-based superconductor. The highest Tc reaches up to 56 K by rare-earth substitution.[4] 2, "122", represented by Co or K-doped $BaFe_2As_2$, is the most studied system due to available high-quality crystals and tunable doping levels.[5] 3, "111" refers to LiFeAs with the maximum Tc of 18 K.[6] 4, "11" is Fe(Se,Te), the prototype of iron-based superconductor without poisonous element As.[7] 5, "21311" is isostructural to 122 system with a divalent group instead of Ba.[8] 6, $A_{0.8}Fe_{1.6}Fe_2$ (A=K, Tl, Cs…), which has ordered Fe-vacancies was discovered recently with $T_c$~30K.[9] The vacancy ordering which is crucial to its superconductivity has been a hot issue in recent research. Due to the difference in structure, the superconducting compounds exhibit various bulk properties.

[*] qmzhang@ruc.edu.cn

Nevertheless, they have many fundamental aspects in common. First, they have similar FeAs or FeSe layers in which superconductivity occurs. It is analogous to copper-oxide planes in cuprates. Second, magnetic order plays a key role in the materials. There is no consensus yet about the relationship between superconductivity and magnetic order. Third, both magnetic and structural transitions are observed in their parent compounds. But the relation between the two transitions is still controversial. Finally, Fe 3d electrons are involved in superconductivity in all the Fe-based superconductors and they have similar Fermi surfaces. More details can be found in the review.[10] In the following we will in turn focus on excitations of phonons, magnons and electrons and some related issues in iron-based superconductor.

2. **General structures of iron-based superconductors and phononic Raman scattering**

2.1. *Crystal structures and Raman phonons*

The six categories of iron-based superconductors are shown in Fig. 1. They have analogous FeAs/FeSe layers as building blocks. Phononic Raman scattering has been carried out for each structure. Among the structures, FeSe ("11") has space group P4/nmm and BaFe$_2$As$_2$ ("122") I4/mmm. Optical phonon modes at Gamma point for both systems are $A_{1g}+2A_{2u}+B_{1g}+2E_g+2E_u$. Raman-active modes include $A_{1g}$ of As/Se, $B_{1g}$ of Fe and two $E_g$ modes of Fe and As/Se.

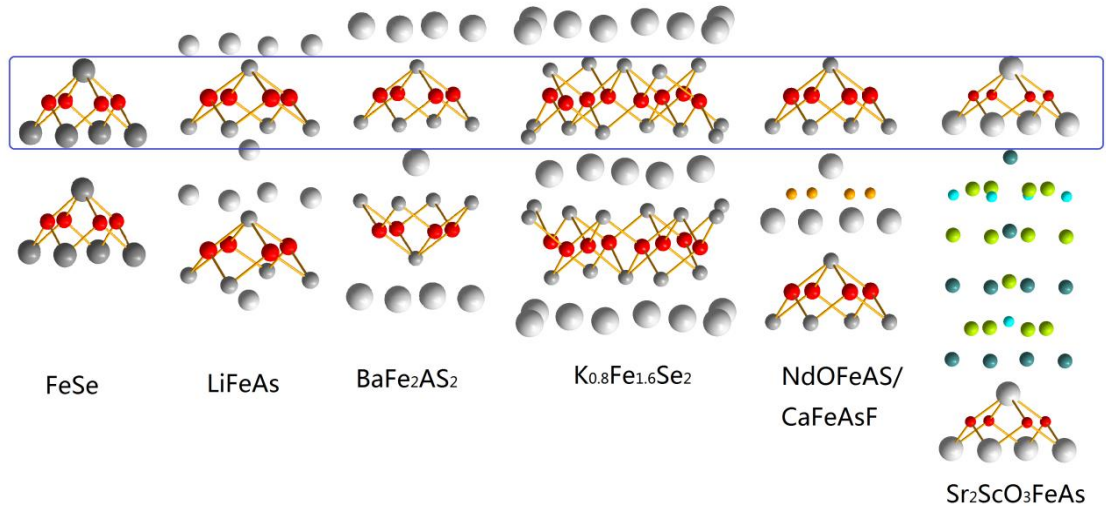

Fig. 1. (Color online) Schematics for six categories of iron-based superconductors. Blue retangular box highlights common FeAs/FeSe layers.

LiFeAs ("111") and LaFeAsOF("1111") also belong to P4/nmm. Besides the above Raman-actives modes, there are additional $A_{1g}$ and $E_g$ of Li or La and $B_{1g}$ and $E_g$ of O because Li and La and O are off-centered.[11-15] Among the Raman-active modes, those related to FeAs/FeSe layers attract most interests due to their possible connections with superconductivity. Fig. 2 shows four Raman-active modes for BaFe$_2$As$_2$. The four modes also appear in the other iron-based systems because BaFe$_2$As$_2$ has a higher symmetry. In 1111 system, Raman phonons exhibit no anomaly around structural, spin-density-wave (SDW) and superconducting transitions. This may suggest a slight structural change and a weak connection between phonons and SDW/superconductivity.[16,17] Similarly, in LiFeAs ("111") without structural and magnetic transitions, both positions and widths of phonons follow a normal anharmonic trend, which implies a weak electron- and spin-phonon coupling.[11] However, anomalies are observed crossing the transitions in 11 and 122 systems, indicating a coupling between phonons and other degrees of freedom. We will discuss it in more details in the following.

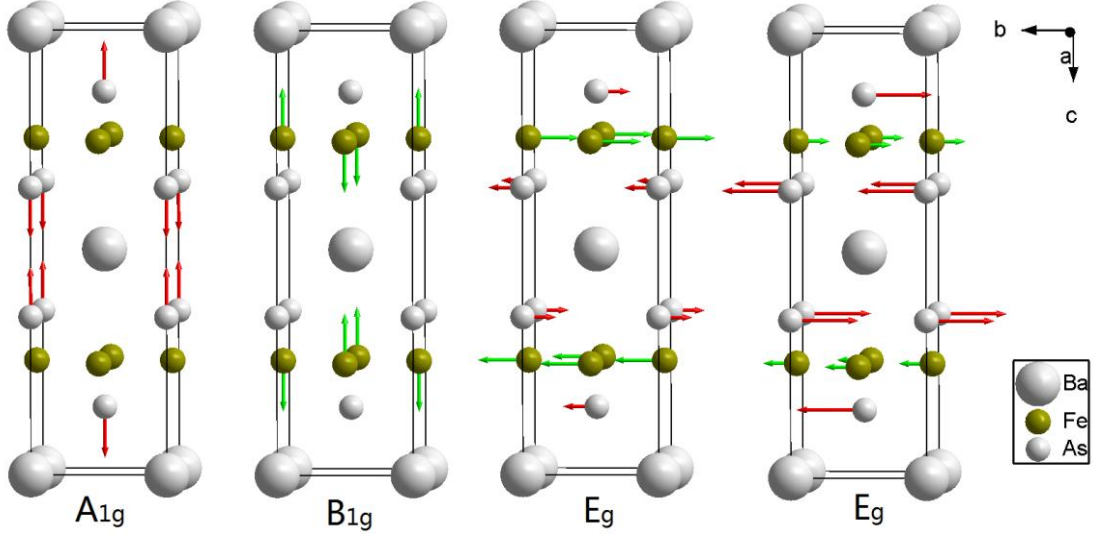

Fig. 2. (Color online) Four Raman-active modes of 122 system.

## 2.2. *Electron-phonon and spin-phonon coupling in 11 and 122 systems*

Some Raman results in 122 system are still controversial. Litvinchuk *et al.*[13] found phonon behavior in 122 $Sr_{1-x}K_xFe_2As_2$ (x=0, 0.4) is similar to that of 1111, i.e., no phonon anomaly exists around structural, magnetic and superconducting transitions, while K. Y. Choi *et al.*[18] reported a jump of $B_{1g}$ mode in frequency and anomalous changes of $A_{1g}$ phonon intensities, which is attributed to SDW gap opening. The anomalies are also seen in parent compound $BaFe_2As_2$.[19] M. Rahcengbeck et al. came to a similar conclusion by Raman scattering measurements in $R_{1-x}K_xFe_2As_2$ (R=Ba, Sr).[20] In addition, they observed an anomalous hardening of $A_{1g}$ mode crossing superconducting transition, which is also found in $B_{1g}$ mode, as reported in $Pr_{0.12}Ca_{0.88}Fe_2As_2$[21] and $Sr_{0.85}K_{0.05}Fe_2As_2$.[22] The hardening is considered to be induced by renomalization of phonon self-energy due to electron pairing.

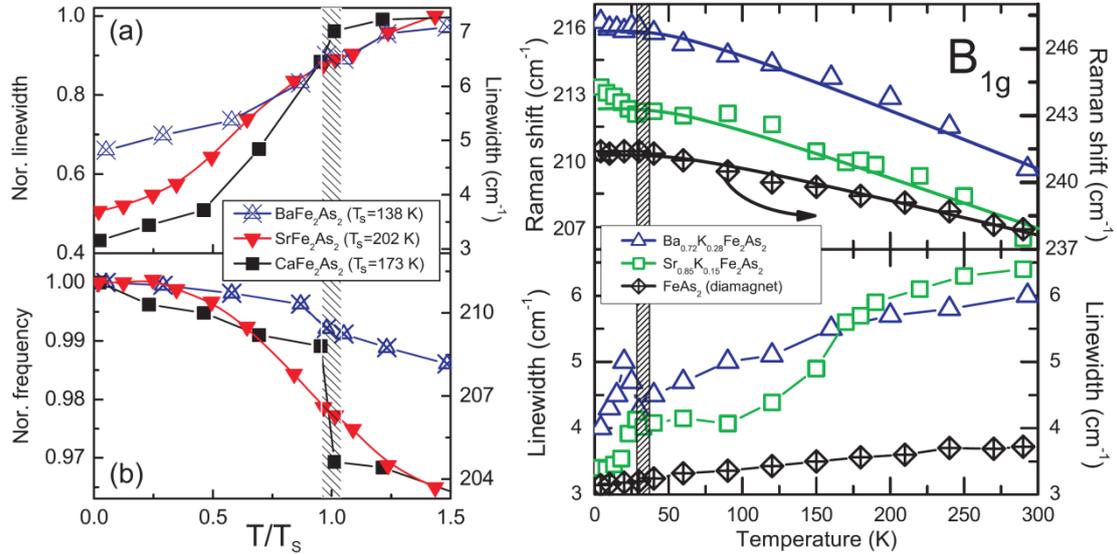

Fig. 3. Left: (Color online) Temperature dependence of B1g phonon frequencies in three 122 parent compounds. Right: Temperature dependence of B1g phonon frequencies in 122 superconducting compounds. The solid lines are fitting curves based on anharmonic effect.[22]

In the left panel of Fig. 3, one can see that the widths of $B_{1g}$ phonons in three parent compounds show obvious slope changes at SDW and structural transitions. This is attributed to phonon self-energy effect due to SDW gap opening. Compared to $BaFe_2As_2$ and $SrFe_2As_2$, $B_{1g}$ mode in $CaFe_2As_2$ demonstrates a larger jump in frequency, which may be associated with the large jump of its lattice parameters. The temperature dependence of $B_{1g}$ mode in superconducting crystals is shown in the right panel of Fig. 3. Phonon frequencies in both crystals deviate from a standard anharmonic model below $T_c$ and increase with decreasing temperatures. The widths also follow the similar evolution and unusual widths can be obtained below $T_c$ if SDW contributions are subtracted. The observation is very similar to the case of cuprate superconductors, where it is due to the opening of $2\Delta$ gap with a size smaller than phonon frequencies.[23] Therefore, the anomalous behaviors at SDW and superconducting transitions are tightly related to electron-phonon coupling. One can estimate electron-phonon coupling constant $\lambda$ from Raman measurements. Basically we have two ways to obtain $\lambda$. One is to use Allen formula $\Gamma = 2\pi\lambda N(0)\omega^2$, where $\Gamma$ is phonon width (~2 cm$^{-1}$ in the current case), $N(0)$ and $\omega$ are density of states at Fermi surface and phonon frequencies, respectively. It gives a coupling constant $\lambda_{B_{1g}}$~0.02. The other way is to estimate $\lambda$ by means of phonon self-energy renormalization induced by electron pairing. The corresponding formula is $\lambda = -\kappa \cdot Re[(\sin u)/u]$,[24] where $\kappa = \omega^{SC}/\omega^N$, $u = \pi + 2i \cdot \cosh^{-1}(\omega^N/2\Delta)$. Then we will have $\lambda_{B_{1g}}$~0.01. The values of coupling constant obtained from both ways are much smaller than theoretically predicted $\lambda$~0.2.[25] This implies that electron-phonon coupling may play a minor role in superconducting pairing.

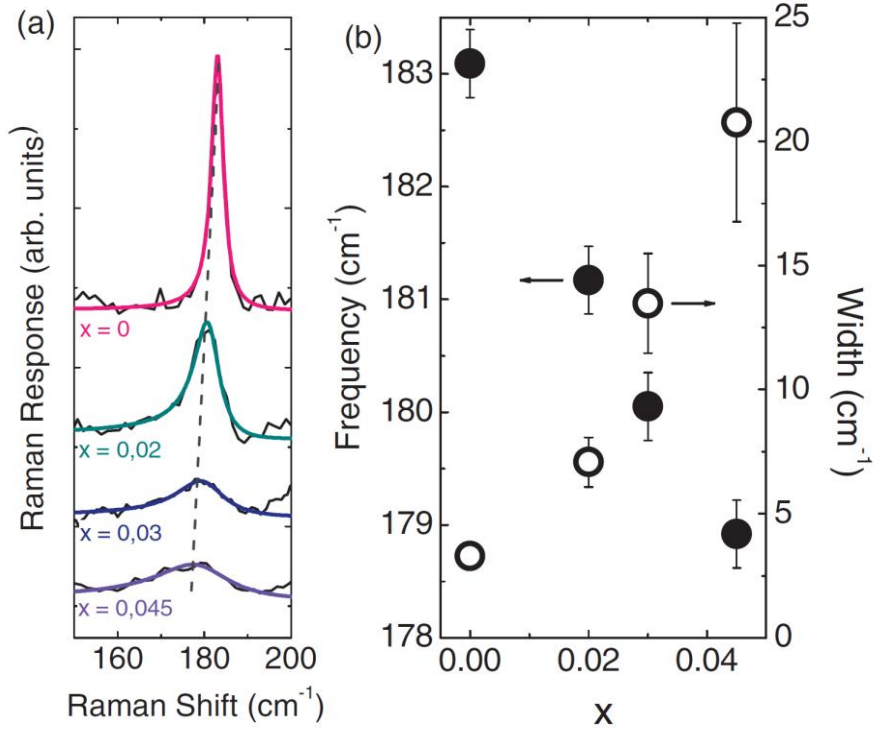

Fig. 4. (Color online) (a) $A_{1g}$ Raman spectra at low temperatures in $Ba(Fe_{1-x}Co_x)_2As_2$. The solid lines are Fano fitting curves. (b) The evolution of $A_{1g}$ phonon frequencies and widths with doping.[26]

Asymmetrical $A_{1g}$ phonon line shape was also reported in $Ba(Fe_{1-x}Co_x)_2As_2$ as shown in Fig. 4,[26] which is also an indication of electron-phonon coupling. And the asymmetry becomes more obvious with increasing doping x. From x=0 to 0.45, phonon frequencies are reduced by 3.5 cm$^{-1}$ and widths increase by 7 cm$^{-1}$. The large change can not be explained simply with a 3% shift of c axis. In point of view of electron-phonon coupling, the phonon softening and widening may be associated with the raising of density of states at Fermi surface. However, the possibilities of spin-phonon or other interactions are not really excluded.

No phonon anomaly at Tc has been reported in 11 system, just as in 111 and 1111 systems. However, at magnetic and structural transitions $A_{1g}$ and $B_{1g}$ modes exhibit unusual behaviors.[27,28]

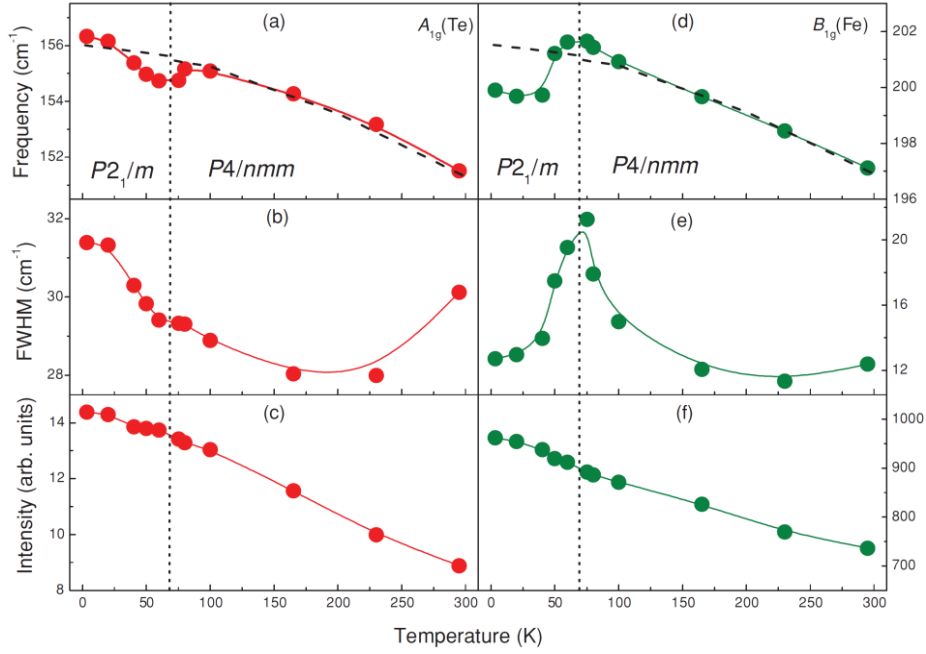

Fig. 5. (Color online) Temperature dependence of frequencies, widths and intensities of A1g and B1g modes. The vertical dashed lines denote transition temperatures.[27]

Fig. 5 shows the temperature dependence of the two modes in frequency, width and intensity in parent compound $Fe_{1.05}Te$. As temperature decreases towards magnetic and structural transitions, $A_{1g}$ mode first decreases and then continuously increases below transition temperatures, while $B_{1g}$ mode shows an opposite evolution. The anomalies are considered to be related to spin ordering because they can not be explained with the smooth evolution of volume of unit cell around transition. Assuming that the main contributions to spin-phonon coupling come from the variations of exchange integration due to the change of bond angle, one can estimate nearest-neighboring and next- nearest-neighboring coupling constants to be $\lambda_{nn} = 3.6 \times 10^{-6}$ meV/(Å$\mu_B$)$^2$ and $\lambda_{nnn} = 1.9 \times 10^{-6}$ meV/(Å$\mu_B$)$^2$, respectively. It should be noted that different bond lengths may modify the momentum of Fe spin. For $B_{1g}$ vibration, two bonds are elongated and the other two shortened. For $A_{1g}$ mode, the lengths of all the four bonds have an in-phase change. This will make an additional contribution to spin-phonon coupling in $A_{1g}$ channel. Moreover, it may induce a spin-orbit frustration as the momentum of Fe spin and orbital occupation are tightly coupled. This may explain the continuous increase of A1g phonon widths after entering into spin-ordered states.[27]

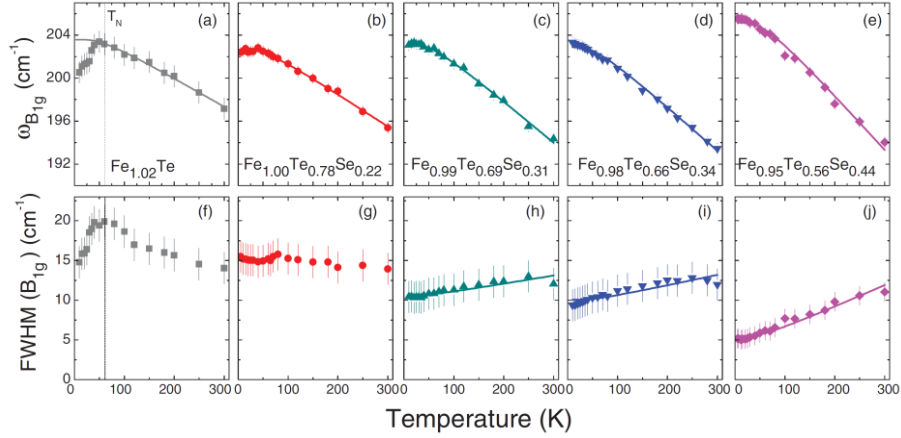

Fig. 6. (Color online) Temperature dependence of B1g phonon frequencies and widths in 11 samples with different doping levels. Solid lines are fitting curves according to anharmonic effect.[28]

Similar anomalies were also reported in Fe(Se, Te),[28] as shown in Fig. 6. With increasing the substitution of Te by Se, the anomalies mentioned above are gradually suppressed. Eventually the phonon modes become conventional anharmonic ones. On the other hand, it seems phonon widths are not affected by disorder induced by Se substitution. Spin-phonon coupling may provide a plausible explanation for it. It was proposed that excess Fe in 11 parent compounds can produce low-energy magnetic fluctuations,[29,30] which may cause the broadening of $B_{1g}$ mode. Anyway, it seems that spin-phonon coupling exists in 11 system and may be correlated with orbital ordering.

In short summary, anomalous phonon changes in both frequency and width crossing SDW/structural transitions are observed in 122 and 11 systems. A stronger spin-phonon coupling may exist in 11 compounds, which is consistent with its larger spin momentum (~$2.5\mu_B$/Fe). Electron-phonon coupling or phonon self-energy effect brought by superconductivity seems weak, which may lie in the too large mismatch in energy between superconducting gap and phonon frequencies.

### 2.3. *Some Raman results in Fe-deficient FeSe-122 system*

The Raman results are much more complicated in the newly discovered Fe-deficient FeSe-based 122 superconductors. A. M. Zhang *et al.* reported that at least thirteen Raman-active modes are observed in $K_xFe_{2-y}Se_2$ crystals, far more than four modes expected for a standard $BaFe_2As_2$-like structure.[31] By delicate symmetry analysis, they deduced a $\sqrt{5}\times\sqrt{5}$ Fe-vacancy ordering pattern, which is in good agreement with neutron scattering experiments performed on the same batch of crystals.[32] Interestingly, a jump in frequency occurs at Tc for $A_g$ mode at 180cm$^{-1}$, implying a particular connection with superconductivity. Further Raman studies on K-substituted crystals and non- superconducting $KFe_{1.5}Se_2$ reveal that the substitution of K by Tl or Rb has little effect on the frequencies of phonons from FeSe layers, consistent with their unchanged $T_c$s.[33] But phonon frequencies and line shapes in non-superconducting sample changed obviously when compared to the superconducting one, which suggests Fe deficiency may be crucial to both electrons and phonons.

A. M. Zhang *et al.* further reported a drop at 44 K in resistivity and susceptibility in $K_xFe_{2-y}Se_2$ crystals.[34] The crystals include two sets of Raman "fingerprints". One is well consistent with $\sqrt{5}\times\sqrt{5}$ structure with $T_c$ of 30 K. The other one comes from fully occupied FeSe layers. In combination of scanning electron microscopy, X-ray diffraction and electron spin resonance, they identified the anomaly corresponds to a 44K-superconducting phase. And its structure is analogous to 122 with excess Fe in the interstitial sites of Se layers. So far, no obvious sign of electron-phonon coupling has been seen in the system.

**2.4. *Isotope effect***

Isotope effect is a key experiment in determining the role phonon plays in pairing. Early isotopic experiments were carried out in SmFeAsO$_{1-x}$F$_x$ (x=0, 0.15) and Ba$_{1-x}$K$_x$Fe$_2$As$_2$ (x=0, 0.4),[35] with $^{16}$O replaced by $^{18}$O and $^{56}$Fe by $^{54}$Fe. The frequency shifts of ~4% and 1.7% for E$_g$ and B$_{1g}$ modes respectively, which means that isotopic replacement is successful. The replacement of $^{16}$O by $^{18}$O has little influence on SDW and superconducting transition temperatures, while the isotopic substitution of Fe drives a shift of the temperatures. The coupling constant α ~ 0.35 implies electron-phonon coupling alone is insufficient to understand pairing mechanism.

However, Granath *et al.* suggested that the observed temperature dependence of phonon frequencies in 1111 can be explained by phonon-phonon rather than electron-phonon interaction, which gives a very small electron-phonon coupling constant α ~ 0.06.[36] P. M. Shirage et al. carried out isotopic experiments on Ba$_{1-x}$K$_x$Fe$_2$As$_2$ with T$_c$~38 K,[37] from which a surprising result α ~ -0.18 was deduced. The reason for the difference is still unclear.

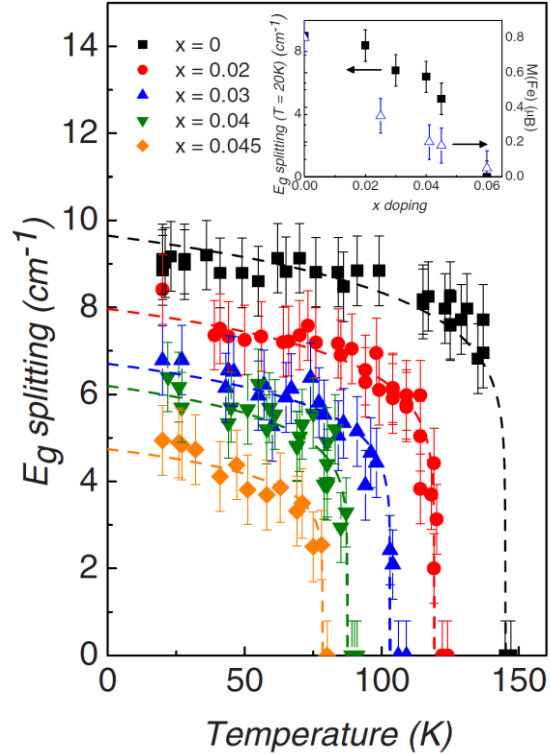

Fig. 7. (Color online) Temperature dependence of Eg splitting at different doping levels in 122 compounds. Inset: Comparison between splitting and spin momentum at 20 K.[19]

**2.5. *E$_g$ phonon splitting***

As we can see in Fig. 2, E$_g$ mode vibrates in-plane and is doubly-degenerate due to identical a and b axis in tetragonal phase. Normally the structural transition from tetragonal to orthorhombic phase is accompanied by E$_g$ splitting into B$_{2g}$ and B$_{3g}$ modes. In 122 system, the splitting would be very small if it is controlled only by the structural transition, as the distortion is really tiny.[38] However, Raman measurements in underdoped Ba(Fe$_{1-x}$Co$_x$)$_2$As$_2$ demonstrate that the splitting up to 10 cm$^{-1}$ is unexpectedly large.[19] In Fig. 7, one can see that the splitting Δω(T) below structural transition temperature just behaves like an order parameter, which follows a power law $\Delta\omega(T) = \Delta\omega(T=0K)\times(1-T/T_s)^\beta$, where $T_s$ is transition temperature and $\beta \sim 0.12$. This suggests a continuous transition if the splitting is regarded as order parameter. It seems that the splitting and Fe spin momentum share a similar evolution with doping. It has been proposed that the large splitting is dominated by spin-lattice interaction rather than structural transition.

At the same time, Raman studies in 122 parent compounds $ReFe_2As_2$ (Re=Ca, Sr, Ba) indicate that the maximum splitting appears in SrFe2As2 and minimum in CaFe2As2.[38] It does not follow the change of spin momentum but is proportional to SDW transition temperature. It was speculated that orbital ordering may be associated with the splitting. More experimental and theoretical work is required to clarify the issue.

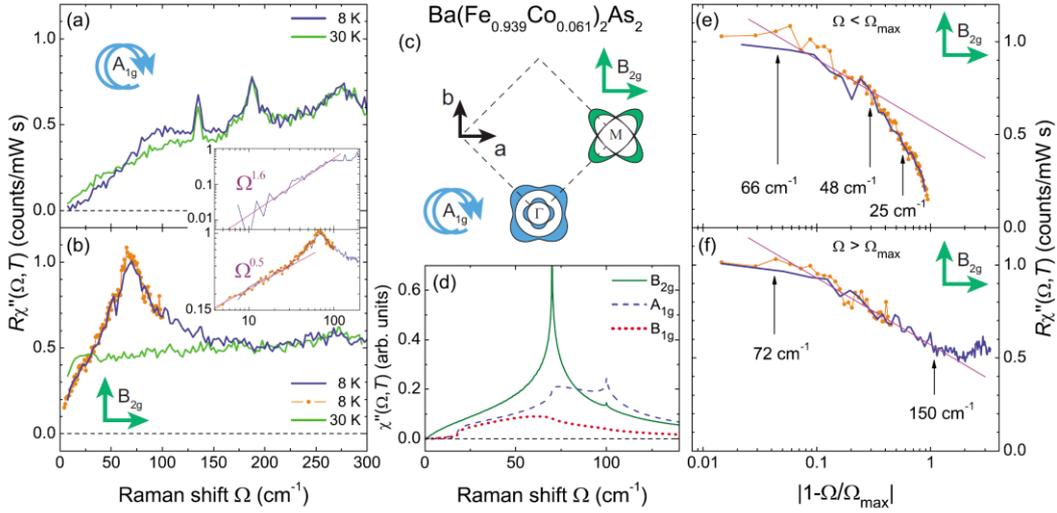

Fig. 8. (Color online) Electronic Raman response in Ba(Fe0.939Co0.061)2As2 in (a) A1g and (b) B2g channels. Gap functions in (c) are used in the calculations (d). Low-energy fitting of B1g spectra.[39]

## 3. Electronic Raman scattering

### 3.1. *Superconducting gap symmetry*

Electronic Raman scattering is one of the most fundamental techniques in exploring superconducting pairing symmetry. Generally high-quality crystals and surfaces are required in electronic Raman scattering experiments, as the signal is weak and easily contaminated by stray light or other background scattering. Some results in electronic Raman scattering have been reported in iron-based superconductors. Fig. 8 shows electronic Raman scattering spectra in $A_{1g}$ and $B_{2g}$ channels in $Ba(Fe_{0.939}Co_{0.06})_2As_2$ measured by B. Muschler et al.[39] Symmetry analysis demonstrates that the spectra in the two channels probe electronic excitations around hole-like Fermi pockets at Γ and electron-like Fermi pockets at M point, respectively. The corresponding pair-breaking peaks are located at 100 cm$^{-1}$ and 70 cm$^{-1}$. Both low-energy parts below the peaks follow power-law behaviors. The deviation begins at 30 cm$^{-1}$ in $A_{1g}$ channel and no deviation is found even down to 5 cm$^{-1}$ in $B_{2g}$ channel. This implies no gap node at Γ pockets and possible nodes at M pockets. Based on the results, electronic Raman response was calculated, and it seems the calculations are in good agreement with experiments. The maximum gaps at hole-like Γ and electron-like M pockets, are determined to be 6.25 and 4.5 meV, respectively.

S. Sugai et al. obtained similar spectra in $Ba(Fe_{1-x}Co_x)_2As_2$, but with lower intensities.[40] They calculated Raman vertexes with real band parameters and attributed both $A_{1g}$ and $B_{2g}$ spectra to hole-like Fermi pockets. The $B_{2g}$ peak at 75 cm$^{-1}$ was assigned as resonance effect. The understanding is consistent with s± scenario.

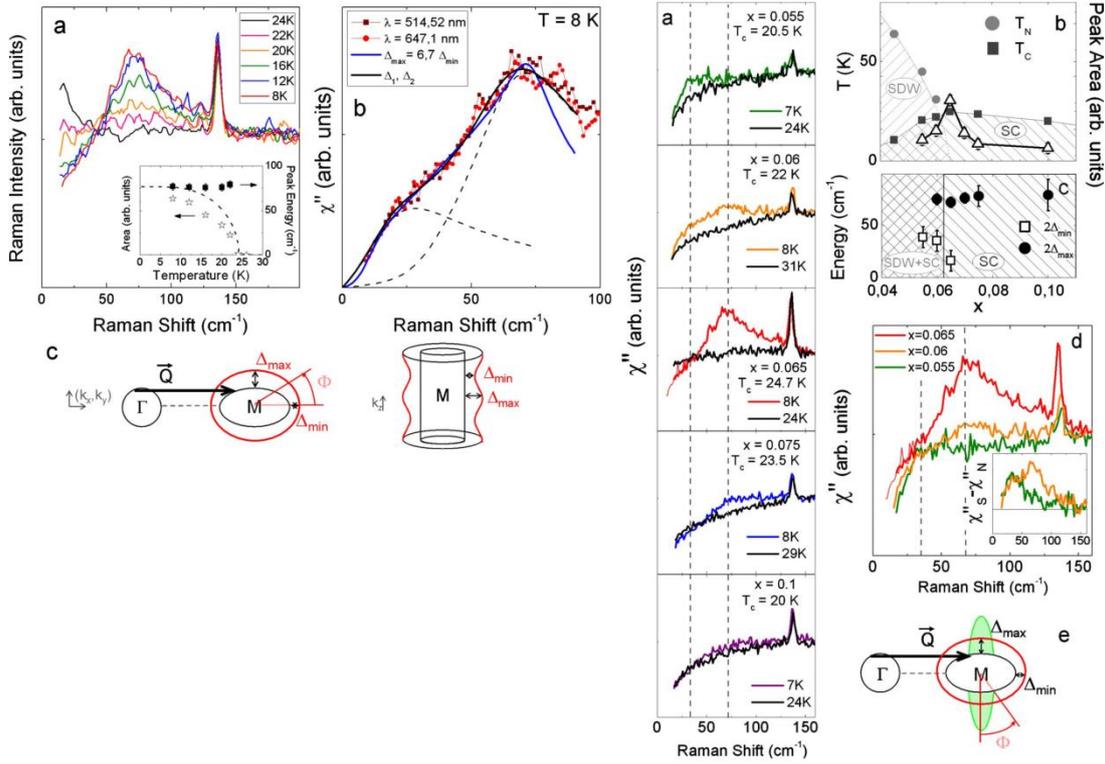

Fig. 9. (Color online) Left: (a) Electronic Raman spectra in B2g channel. Inset: Temperature dependence of positions of pair-breaking peaks. (b) Two fitting results: blue and black line correspond to one and two-gap models, respectively. Right: (a) B2g spectra with different doping levels. (b) Phase diagram on basis of integrated intensities of pair-breaking peaks. (d) Comparison of B1g spectra between different samples. (e) Schematic of competition between SDW and superconductivity at M points.[41]

Electronic Raman scattering studies on Ba(Fe$_{1-x}$Co$_x$)$_2$As$_2$ were also performed by L. Chauvière et al.[41] They measured $B_{2g}$ spectra which is similar to those in Ref. 39 and considered the spectra are contributed by excitations around electron-like M pockets.(Fig. 9) They fitted the data with two scenarios: standard BCS model with broadening parameter Γ and anisotropic s-wave with two gaps. Both fitting results are in good agreement with experiments, so the evidence is insufficient to conclude there exists only one gap or two gaps in the system. In particular, the B2g pair-breaking peak is rapidly suppressed with doping. This was considered to be associated with SDW transition rather than disorder, as Tc and residual resistivity exhibit only a weak doping dependence. They proposed that there might exist a strong competition between superconductivity and SDW and hence a partial nesting in SDW state would significantly affect successive superconducting pairing.

K. Okazaki et al. carried out electronic Raman scattering measurements in 11 system Fe$_{1+y}$Te$_{1-x}$Se$_x$.[42] Calculations indicate that $A_{1g}$ spectra correspond to hole-like Fermi surfaces around Γ points. Unfortunately, only a slope-like feature rather than a peak is observed due to very weak signal. Assuming that the feature originates from pair-breaking process, one can get a gap of ~5 meV.

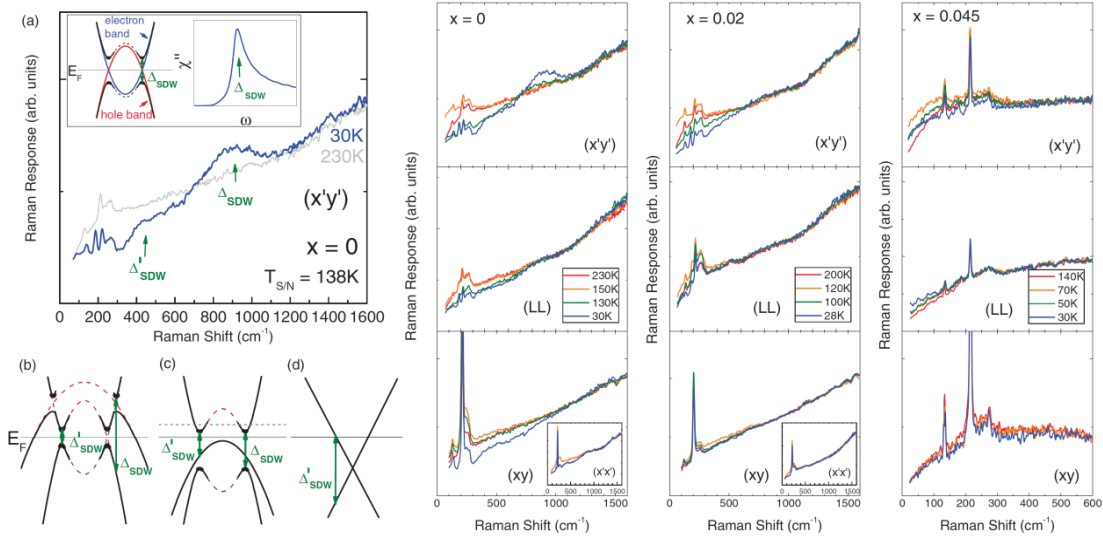

Fig. 10. (Color online) Left: (a) Raman response in BaFe2As2 at different temperatures. Inset: Schematic of SDW gap opening and Raman response under perfect nesting. (b) Band crossing of one electron band and two hole bands. Right: Raman spectra under different scattering configurations. Yellow and green lines refer to measurement temperature close to transition temperature.[26]

### 3.2. *SDW gap*

Electronic transition between different bands can also be observed in Raman spectra. Below SDW transition temperature, a feature near 900 cm$^{-1}$ and a suppression of electronic background below 400 cm$^{-1}$ are seen simultaneously in x'y' channel in Ba(Fe$_{1-x}$Co$_x$)$_2$As$_2$, where x'(y') denotes the orientation of Fe-As bonds. (Fig. 10)[26] In contrast, they are absent above the temperature. Assuming that the feature at 900 cm$^{-1}$ is a simple d-band transition, it must be the transition between d$_{x^2-y^2}$ and d$_{z^2}$ orbitals because it appears only in x'y' channel. But this contradicts other experiments and calculations. It was proposed that the transition between folding bands linked to SDW gap might explain the feature. And the spectral suppression below 400 cm$^{-1}$ can be understood as a direct indication of SDW gap opening.

### 4. Two-magnon Raman scattering

Two-magnon Raman scattering can give some important spin-related information such as exchange coupling constant, evolution of spin order with magnetic fields or temperatures etc. In cuprates, Raman scattering can precisely determine exchange energy due to single parameter and isotropy in plane. However, it is a bit complicated in iron-based superconductors because both ferromagnetic coupling between nearest-neighboring Fe spins and antiferromagnetic coupling between next-nearest-neighboring spins bridged by As, need to be considered simultaneously.

### 4.1. *Two-magnons in FeAs-based compounds and Fe(Se, Te)*

Two-magnon peak in Ba-122 was reported by S. Sugai *et al.*[43] A broad peak around 2200 cm$^{-1}$ at 10 K is still visible above SDW transition temperature and shifts to 2600 cm$^{-1}$ at room temperature. This was regarded as a sign of short-range spin correlations. The similar feature was also observed in FeTe$_{1.074}$ and FeTe$_{0.6}$Se$_{0.4}$, peaked at 2300 cm$^{-1}$. With a classical Heisenberg model, two-magnon energy can be written as: $4S(J_{1a} - J_{1b} + J_{2a} - J_{2b}) - J_{2a}$, where $J_{1a(b)}$ and $J_{2a(b)}$ refer to exchange coupling between nearest- (next-nearest-) neighboring spins along *a (b)* axis. With the values given by neutron scattering $SJ_{1a}$ = -7.6 meV, $SJ_{1b}$ = -26.5 meV, $SJ_{2a}$ = 46.5 meV, and $SJ_{1a}$ = -34.9 meV, two-magnon energy is estimated to be ~ 354.7 meV, slightly larger than experimental one.[32] The clear observation of two-magnon peaks at temperatures much higher than SDW transition, implies a strong short-range spin fluctuations. This is apparently different from cuprates, the reason may lie in the competitive exchange interactions and itinerant magnetism.

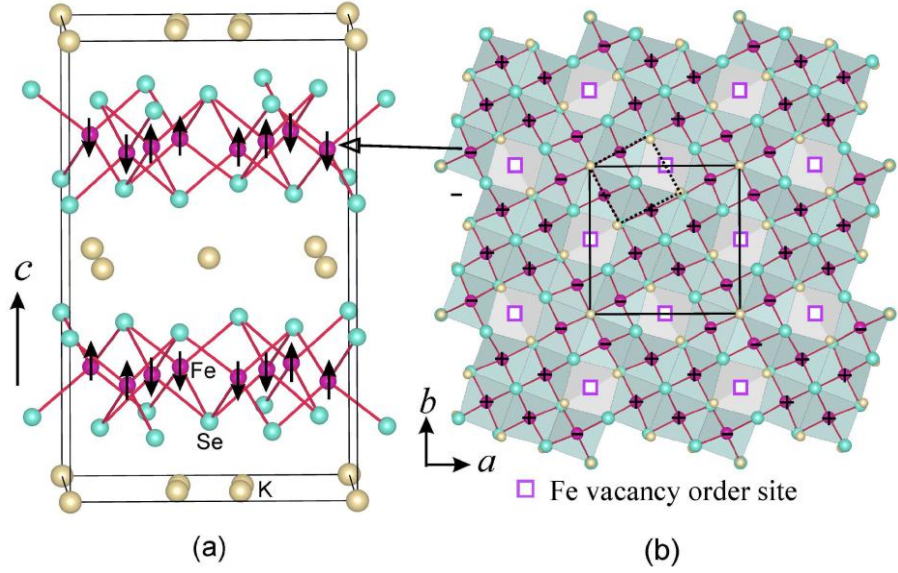

Fig. 11. (Color online) Crystal and magnetic structures of Fe-deficient KxFe2-ySe2.[32]

**4.2. Interplay between antiferromagnetism and superconductivity in Fe-deficient KFeSe system**

Newly discovered Fe-deficient $K_xFe_{2-y}Se_2$ is distinct from other Fe-based superconductors in some aspects. First, √5×√5 pattern of Fe-vacancies reduces crystal symmetry from $D_{4h}$ to $C_{4h}$, as shown in Fig. 11.[32] Second, four nearest Fe spins form a spin cluster and spin clusters are antiferromagnetically coupled. Finally, it has a very high magnetic transition temperature of ~500 K and a large spin momentum of 3.3$\mu_B$/Fe in spin-ordered state. Below 30 K, it enters into superconducting state. Then a key issue on relationship between magnetism and superconductivity is raised immediately. Many experiments are tend to be explained with macro-phase separation, i.e., completely different phases are respectively responsible for superconductivity and antiferromagnetism. Two-magnon Raman measurements in the system can provide further information on the issue.

Two-magnon peaks in three superconducting crystals are clearly shown in Fig. 12, which are located around 1600 cm$^{-1}$.[44] In a local moment Heisenberg picture, two-magnon energies for the spin-cluster configuration are *2S(-2J$_1$+J$_1$'-J$_2$+2J$_2$'+J$_3$)-SJ$_1$'* or *2S(-2J$_1$+J$_1$'-J$_2$+2J$_2$'+J$_3$)-SJ$_2$'*, where $J_1$, $J_2$ and $J_3$ are nearest-, next-nearest- and third-nearest-neighboring exchange energies and prime denotes inter-cluster spin exchange. Neutron scattering gives $SJ_1$ = -36±2, $SJ_1$' = 15±8, $SJ_2$ = 12±2, $SJ_2$' = 16±5, and $SJ_1$ = 9±5 meV. Then we can estimate the two-magnon energy to be ~216 meV, in good agreement with experimental values.

Two-magnon peak becomes more narrower and its intensities continuously increase with lowering temperatures. Most interestingly, a reverse of intensities occurs when crossing $T_c$. The intensities rapidly drop down below $T_c$, which indicates a strong interplay between magnetism and superconductivity. A direct explanation is the coexistence of magnetism and superconductors but we need to face other difficulties such as insulating parent compound with a large gap, very high Neel temperature etc. In a phase separation scenario, the clear interplay implies a micro- rather than macro-phase separation because proximity effect in a macroscopically separated sample cannot bring a

5-10% intensity drop of two-magnon peak.

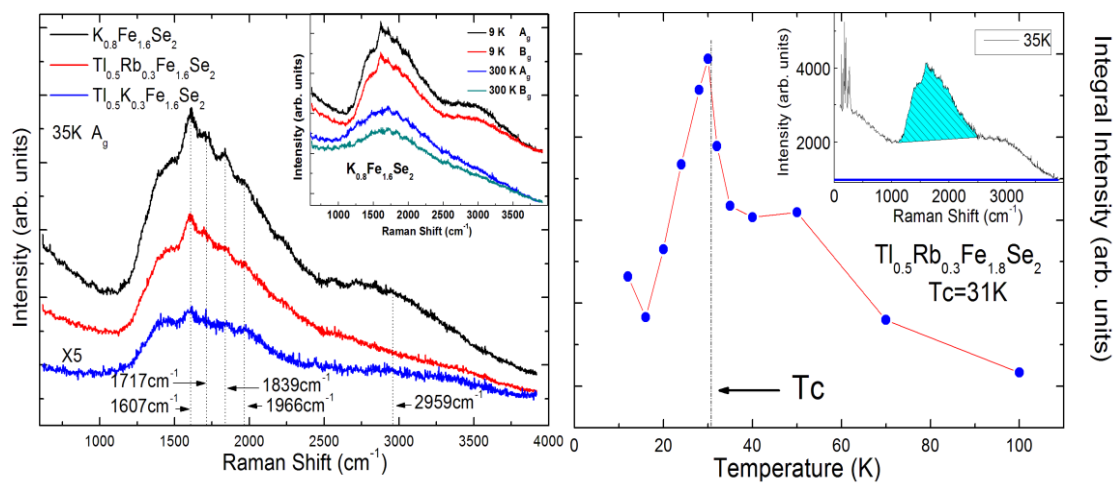

Fig. 12. (Color online) Left: Two-magonon peaks in three superconducting $K_xFe_{2-y}Se_2$ crystals. Right: Temperature dependence of intensities of two-magnon peaks.[44]

## 5. Summary

Raman scattering has been extensively used to study phonons, magnons and electronic pairing in iron-based superconductors. It substantially contributes to our understanding of structural changes, coupling between various excitations, paring symmetry and mechanism, anisotropy, and competition between superconducting and other electronic states etc. On the other hand, most iron-based superconductors and their parent compounds are typical alloys with good conductivity, which brings a partial screening on incident light and hence weak Raman signals. That may cause some contradicting Raman results. It can be expected that with further improvement of crystal quality of different iron-based systems, more consensuses can be reached. And Raman scattering will continue to play a crucial role in exploring iron-based and other correlated electron materials.

**Acknowledgements**

This work was supported by the NSF of China (Grant Nos. 11174367&11034012), the 973 program (Grant Nos. 2011CBA00112 & 2012CB921701), the Fundamental Research Funds for the Central Universities, the Research Funds of Renmin University of China.